# Magnetic Iron Nanocubes Effectively Capture Epithelial and Mesenchymal Cancer Cells

Dhananjay Suresh,[⊥] Shreya Ghoshdastidar,[⊥] Abilash Gangula, Soumavo Mukherjee, Anandhi Upendran, and Raghuraman Kannan*

| ACCESS | Metrics & More | Article Recommendations | Supporting Information |

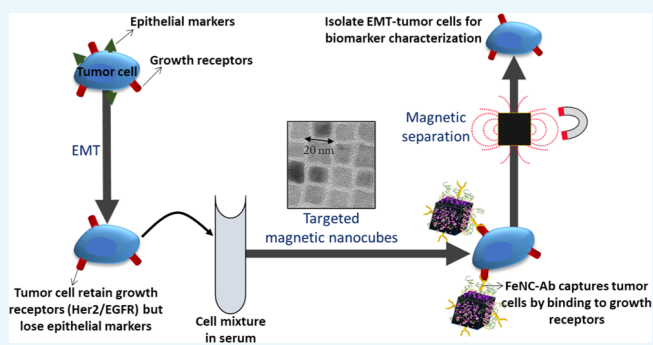

**ABSTRACT:** Current methods for capturing circulating tumor cells (CTCs) are based on the overexpression of cytokeratin (CK) or epithelial cell-adhesion molecule (EpCAM) on cancer cells. However, during the process of metastasis, tumor cells undergo epithelial-to-mesenchymal transition (EMT) that can lead to the loss of CK/EpCAM expression. Therefore, it is vital to develop a capturing technique independent of CK/EpCAM expression on the cancer cell. To develop this technique, it is important to identify common secondary oncogenic markers overexpressed on tumor cells before and after EMT. We analyzed the biomarker expression levels in tumor cells, before and after EMT, and found two common proteins—human epidermal growth factor receptor 2 (Her2) and epidermal growth factor receptor (EGFR) whose levels remained unaffected. So, we synthesized immunomagnetic iron nanocubes covalently conjugated with antibodies of Her2 or EGFR to capture cancer cells irrespective of the EMT status. The nanocubes showed high specificity (6−9-fold) in isolating the cancer cells of interest from a mixture of cells spiked in serum. We characterized the captured cells for identifying their EMT status. Thus, we believe the results presented here would help in the development of novel strategies for capturing both primary and metastatic cancer cells from patients' blood to develop an effective treatment plan.

## ■ INTRODUCTION

Isolation of circulating tumor cells (CTCs) from the blood of cancer patients and analyzing them enables the clinician to predict the disease status, drug resistance, and the selection of appropriate therapy. Food and Drug Administration (FDA)-approved CellSearch is currently used for the detection of CTCs in a variety of metastatic tumor types to predict the overall survival and progression-free survival in patients.[1] This system utilizes magnetic microbeads coated with an antibody (Ab) specific to epithelial cell-adhesion molecule (EpCAM) for the enrichment of CTCs from the patient's blood. Even though these magnetic beads are beneficial and widely used in clinics, the system has several drawbacks.[2,3] Notably, the system detects only EpCAM-positive CTCs and fails to capture tumor cells with no epithelial markers.[4]

The tumor cells lose cytokeratin (CK) and EpCAM while undergoing epithelial-to-mesenchymal transition (EMT), a process that occurs during metastases.[5,6] In fact, the loss of these epithelial markers makes CTCs elastic, aiding cell movement through the extracellular matrix of a tumor leading to metastasis (Figure 1).[7,8] Importantly, the transitioned CTCs with no epithelial markers provide crucial information about the metastasis and effective treatment options.[9−12] Therefore, it is vital to develop a capturing technique independent of CK/ EpCAM expression on the cancer cell. The technique based on microfluidics for sorting cells has been developed for capturing mesenchymal cells.[13] Attempts have been made to increase the efficiency of the microfluidic system by combining with immunomagnetic beads.[14,15] The ideal system to efficiently capture the EMT transitioned cells is still lacking.

One possible way to capture the cells with high efficiency is to develop immunomagnetic beads that are selective, in capturing both epithelial and mesenchymal cancer cells, and powerful by selectively removing the cells from the milieu. To develop a selective immunomagnetic bead, we need to identify the common biomarkers overexpressed on tumor cells before and after EMT. Therefore, we analyzed the biomarker expression levels in tumor cells, before and after EMT, and found two common proteins—human epidermal growth factor receptor 2 (Her2) and epidermal growth factor receptor (EGFR), whose levels remained unaffected. On the other hand, to develop a powerful immunomagnetic bead, we need particles with high magnetic moment. Traditional sphere-shaped magnetic beads have a magnetic moment of 5−40 emu/g.[16−20] The magnetic moment can be increased by decreasing the size of beads or changing the spherical shape to a cube.[17,21,22] For example, nanosized spherical particles (∼40 nm) show magnetic moment >40 emu/g, whereas the cube-shaped nanoparticles of the same size exhibit higher magnetic moment than the spherical counterparts.[22] Therefore, in the present study, we have developed smaller sized nanocubes attached with biomarkers expressed in EMT cells and studied their efficacy in cell capture.

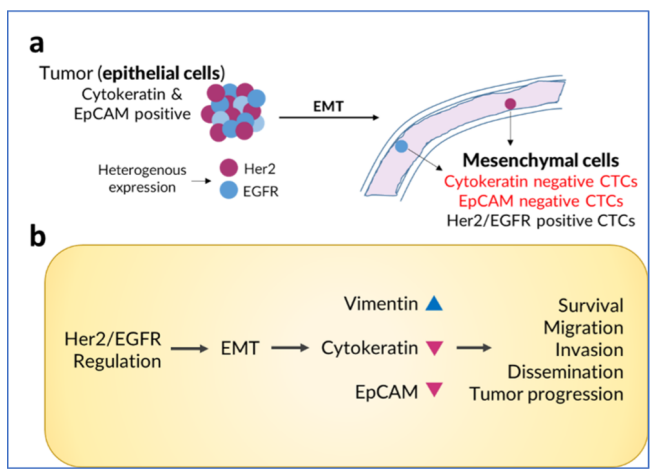

**Figure 1.** (a) Schematic illustration of the migration of tumor cells after undergoing EMT and (b) regulators and markers of the EMT process in tumor cells and metastatic abilities.

Briefly, we followed a two-pronged approach for the isolation of cancer cells. First, we synthesized paramagnetic 20 nm iron oxide nanocubes (FeNCs) with a high magnetic moment of 65 emu/g. Second, we conjugated antibodies to the particles to obtain immunomagnetic iron nanocubes. We chose Her2 (ERBB2) and EGFR (ERBB1) antibodies for functionalizing the nanocubes as they play critical roles in regulating

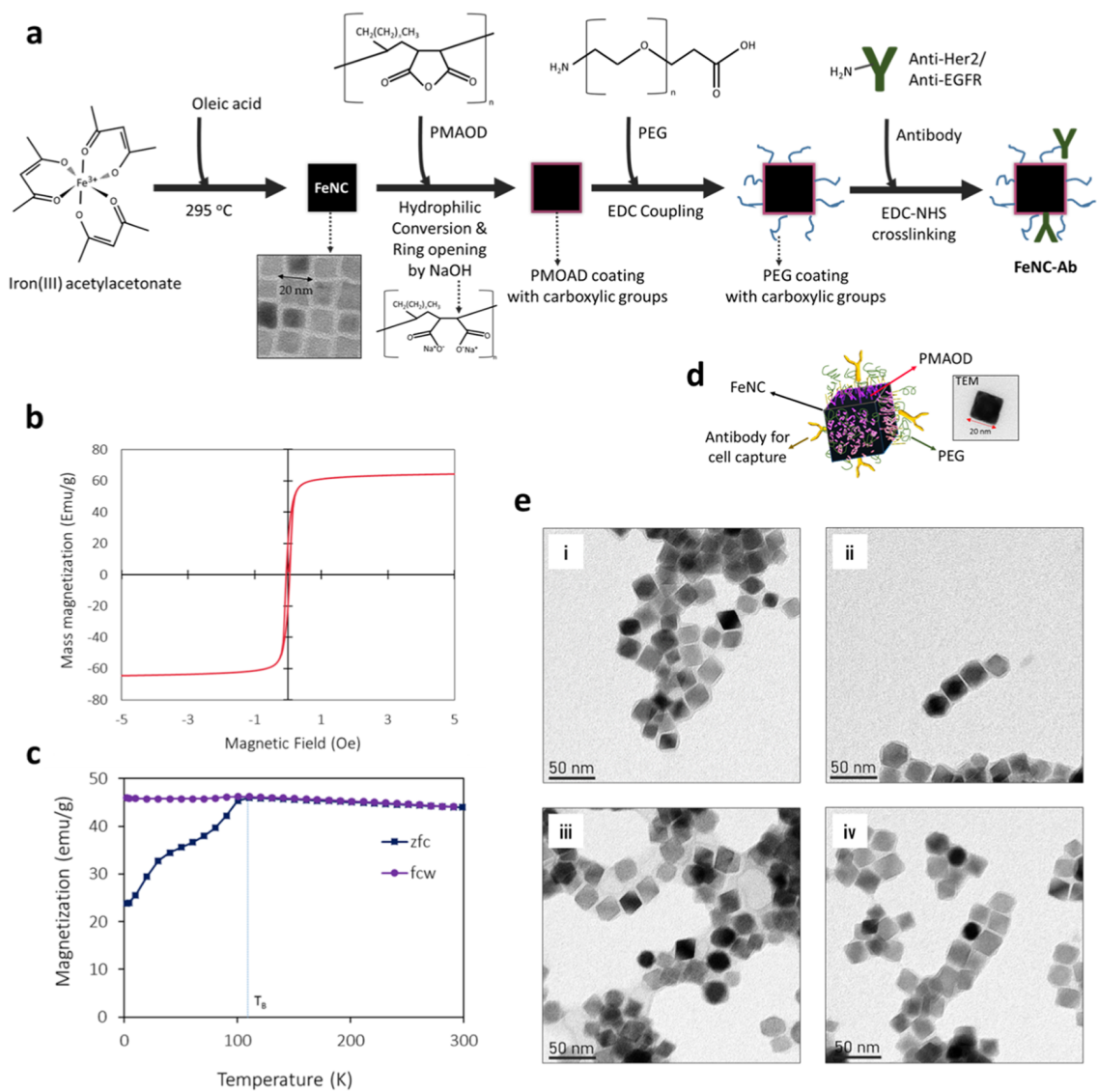

**Figure 2.** (a) Schematic diagram of the preparation of iron oxide nanocube constructs; magnetic properties for FeNC (b) superconducting quantum interference device (SQUID) hysteresis; (c) zero-field curve; (d) schematic illustration of iron nanocubes conjugated to the antibody; and (e) transmission electron microscopy (TEM) images of (i) hydrophilic FeNC, (ii) PEGylated FeNC, (iii) FeNC−Her, and (iv) FeNC−EGF.

EMT.[23,24] It is anticipated that their expression levels often remain unaffected in the cancer cells.[25−28] Using the cube-shaped nanoparticles, we developed a highly efficient immunomagnetic platform, functionalized with antibodies, for the isolation of cancer cells with or without epithelial markers. We avoided the conventional strategy of using gold or silica shell over the magnetic particles, which can potentially decrease the magnetic moment of the final construct.[29,30] Instead, we employed poly(maleic anhydride)-based ring-opening strategy to generate carboxylic functional groups that allow covalent conjugation with biomarker-specific antibodies, as well as increase the hydrophilicity and solubility of the FeNC.

We analyzed the expression levels of biomarkers that are retained and overexpressed on the membrane of cancer cells before and after EMT. The expression levels of Her2 or EGFR were confirmed to remain unaffected, before and after EMT stimulation in cancer cells. Targeting these biomarkers, unaffected by EMT, would allow us to capture cancer cells irrespective of the expression of epithelial antigens.[10−12] Before the EMT process is fully complete, there is an intermediary stage known as partial EMT that may express both EMT regulators such as Her2/EGFR and also epithelial antigens.[31] At this stage, both Her2 and EGFR are activated, resulting in overexpression in the cell surface. However, epithelial antigen levels may fluctuate depending on the environment.[32,33] Therefore, combining an anti-epithelial marker and anti-Her2/EGFR markers with nanocubes may present an efficient approach to capture all CTC subpopulations in a given sample. In this article, we present the results of the following studies: (i) synthesis and characterization of FeNC functionalized with Her2 or EGFR antibodies; (ii) evaluation of the expression levels of biomarkers in cancer cells before and after EMT; (iii) determination of FeNC's ability to capture epithelial and mesenchymal cancer cells; and (iv) evaluation of epithelial markers in captured cells.

## ■ RESULTS AND DISCUSSION

**Synthesis and Characterization of Magnetic Nanocubes.** We synthesized FeNC by thermal decomposition (295 °C) of iron−acetylacetonate complex that generates $Fe^{3+}$ ions that are reduced by polyol to produce nanocrystalline seeds (Figure 2a). By carefully maintaining the growth conditions, we obtained uniform nanocubes with an edge length of 20 nm. The cubic shape imparted a higher magnetic moment of 65 emu/g (Figure 2b,c), relative to previously known spherical nanoparticles, which showed 40−50 emu/g.[21,22] In comparison, magnetic beads of size 0.2−1 mm exhibit a saturation magnetization of 10−25 emu/g.[17] Commercially available Dynabeads show a magnetic moment of 12 emu/g.[18] The magnetic moment data obtained for FeNC synthesized in this study is similar to previously reported iron nanocubes.[34] The synthetic protocol used in this study facilitates the transition of ferromagnetic to paramagnetic FeNC crystals;[34] consequently, the nanoparticles are in a nonmagnetized form in the absence of an externally applied field. This characteristic property enhances the utilization of FeNCs in cell capture and selective isolation. Additionally, nanocubes exhibit high surface area, excellent binding capacity, and higher analytical sensitivity as compared to traditional spherical nanoparticles.[35]

As synthesized, the FeNCs coated with oleic acid are hydrophobic and inefficient in capturing cells under aqueous conditions. Therefore, we modified the surface of the particles with hydrophilic molecules to impart water solubility. Recent studies have utilized amphiphilic molecules such as poly-(maleic anhydride-alt-1-octadecene) (PMAOD) for making nanoparticles water-soluble (Figure 2a).[36,37−40] Importantly, PMAOD adds functional −COOH groups to the nanoparticles for conjugating with antibodies.[36] We replaced the oleic acid on the surface of FeNC with PMAOD and, subsequently, hydrolyzed the rings of maleic anhydride to obtain carboxylic groups (Figure 2a). Further, to prevent the interparticle attraction of FeNCs, which lead to agglomerates, we introduced "poly(ethylene glycol) (PEG)" molecules as a stabilizing linker. The chosen PEG linker "$NH_2$-PEG-COOH" (molecular weight (MW) 2000) increased the interparticle dispersity (Figure 2a,d).[41] Additionally, as the PEG linker contains terminal carboxylic groups, the total available carboxyl groups per nanoparticle remain unchanged before and after PEGylation. We used N-(3-dimethylaminopropyl)-N′-ethyl-carbodiimide hydrochloride (EDC)−N-hydroxysuccinimide (NHS) coupling agent for conjugating PEGylated FeNCs with antibody (Ab) of our choice (Figure 2a,d). It is important to ascertain that the specificity of the antibody is retained after conjugating with PEGylated FeNC. We chose the "antiglobin and globin" to validate the specificity by enzyme-linked immunosorbent assay (ELISA), wherein we conjugated the antibody of globin with PEGylated FeNC and globin as an analyte (Figure S1). The ELISA data showed that the FeNC−Ab−globin construct displayed a linear correlation toward the antigen concentration, confirming that the antibody retains its specificity upon conjugation. Based on the encouraging results, we conjugated FeNC with Her2 or EGFR antibodies for evaluating their ability to capture cancer cells that overexpress the respective receptors.

As a next step, we characterized all of the synthesized nanoparticles using conventional analytical techniques. TEM images of hydrophilic FeNC, PEGylated FeNC, and FeNC−Ab constructs showed the uniform shape of the particles (Figures 2e and S2). FeNC−Ab constructs are stable in aqueous solutions and showed a hydrodynamic size of ∼175 nm with a negative ζ potential (Table 1 and Figure S3).

Table 1. Size and ζ Potential Characteristics for the As-Synthesized Hydrophilic FeNC, PEGylated FeNC, and Antibody-Attached, FeNC−Her and FeNC−EGF

| construct | hydrodynamic size ($d$, nm) | ζ (mV) | TEM edge length (nm) |
|---|---|---|---|
| hydrophilic FeNC | 148 | −63 | 20 |
| PEGylated FeNC | 142 | −5.8 | 20 |
| FeNC−Her | 171 | −6.5 | 20 |
| FeNC−EGF | 175 | −5.8 | 20 |

**Biomarker Selection for Cell Capture.** The overexpressed growth factor receptors, Her2 and EGFR, are considered to be the essential drivers for cell proliferation and regulation of EMT.[23−25,27] To identify if Her2 or EGFR can be used as suitable biomarkers for targeting and isolating cancer cells after EMT, we investigated the expression levels of growth receptors in multiple cancer cells by Western blot (WB) analysis. In this study, we selected three cancer cell lines, namely, A549, HCC827, and MCF-7. We isolated the lysates and performed WB analysis and the results are shown in Figure 3a. The data showed that A549 expresses high levels of both Her2 and EGFR, with Her2 showing 1.6 times higher expression than EGFR; HCC827 expresses high EGFR (3.9 times higher than Her2 relatively) and low Her2, whereas MCF-7 expresses high levels of Her2 and no expression of EGFR. The expression levels agreed with previously published studies.[25−28] Therefore, for Her2-based capturing of cells, we used A549 as a positive and HCC827 as a negative control. On the other hand, for EGFR-based cell capture, we used HCC827 as a positive and MCF-7 as a negative control.

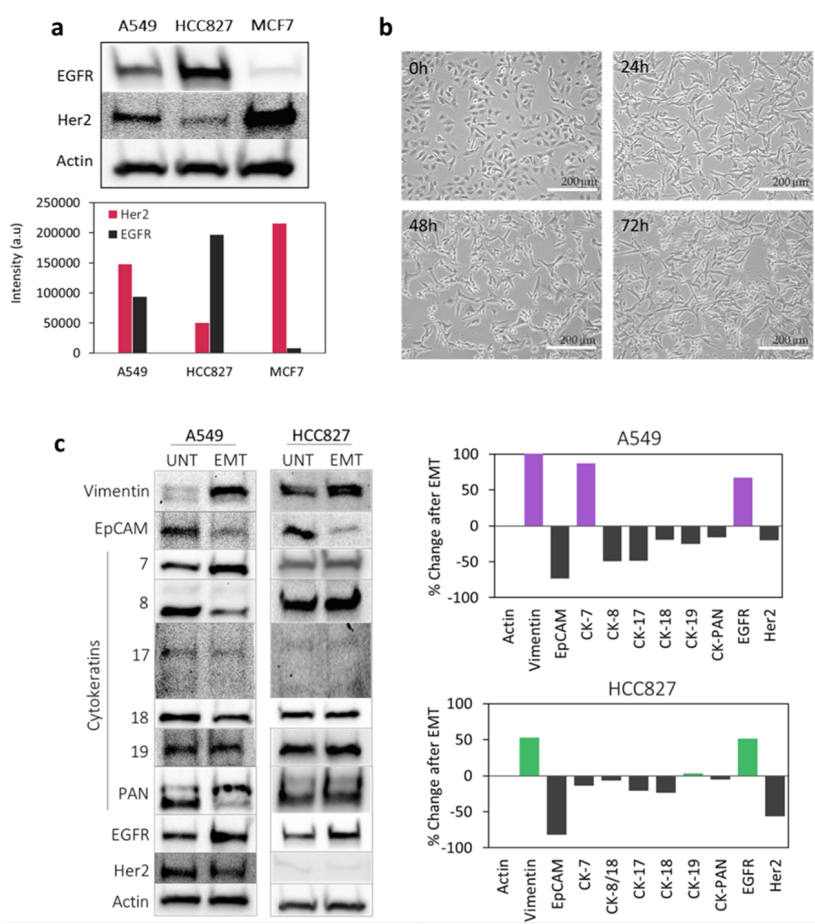

**Figure 3.** (a) Western blot images showing the native expression levels of EGFR and Her2 in A549, HCC827, MCF-7 cells, and its corresponding band densitometry plot. (b) Morphological changes during EMT conversion for 0–72 h in A549 cells. After EMT conversion, the cells appear more elongated in shape compared to untreated (Figure S4). Images were taken using bright-field microscopy at a 10× magnification, and the scale bar represents 200 μm. (c) Western blot images showing changes in the expression levels of proteins in A549 and HCC827 cells before and after induction of EMT, and its corresponding band densitometry plot.

**Expression Levels of Protein Markers in EMT-Stimulated Cancer Cells.** As a next step, we artificially stimulated EMT in our positive target cell lines (A549 and HCC827) to monitor the changes in expression levels of EMT markers, epithelial antigens, Her2, and EGFR proteins. We used TGFβ-1 and EGF to induce EMT in both A549 and HCC827.[42,43] As shown in Figure 3b, within 72 h, the cells lost their epithelial adhesion on the substrate and transformed into an invasive elongated state confirming EMT stimulation. The lysates from these EMT-stimulated cells were then analyzed using WB (Figure 3c). The data showed increased expression levels of vimentin in both the cell lines after the induction of EMT (increase of 171% in A549 and 53% in HCC827). The upregulation of vimentin is characteristic of mesenchymal cells,[44] and the result further validates successful EMT stimulation in cells. As shown in previous studies, we also observed a decrease in expression levels of most cytokeratins (CK; −31% in A549 and −12% in HCC827) and EpCAM epithelial markers (−73% in A549 and −82% in HCC827) in the EMT-induced cells.[5,6,25,27,44] A slight decrease in Her2 expression levels (−20%) in A549 cells was observed after EMT, whereas the EGFR expression levels increased in both A549 and HCC827 EMT cells (67 and 52%, respectively). Based on the data, it is evident that both Her2 and EGFR can serve as biomarkers; therefore, their antibodies can be utilized for capturing A549 and HCC827 cells, respectively, before and after EMT transformation. Indeed, previous studies have also shown that both Her2 and EGFR biomarkers play crucial roles in regulating EMT and also overexpressed during the process.[23−25,27] Thus, targeting these two biomarkers allows the cell capture to be more efficient and independent of their EMT status (CK and EpCAM levels).

**Magnetic Nanocubes for Capturing Epithelial Cancer Cells.** We first investigated the selectivity of the FeNC−Her and FeNC−EGF to capture epithelial cancer cells. As per our WB analysis, A549 showed a high expression of both Her2 and EGFR. Therefore, we hypothesized that the FeNC−Her or FeNC−EGF would be able to capture A549 cells selectively; while HCC827 showed only the EGFR expression, we postulated that only FeNC−EGF would be able to capture HCC827 cells. To test the hypotheses, we prepared a mixture of A549 and HCC827 in equal proportions and independently labeled with different color dyes (for example, A549 with a red membrane dye, while HCC827 with a blue nuclear dye). Subsequently, we used FeNC−Ab constructs to capture cells for isolation by magnetic separation. The cells were then counted by fluorescence and identified by the corresponding color (Figure 4; see the Methods section).

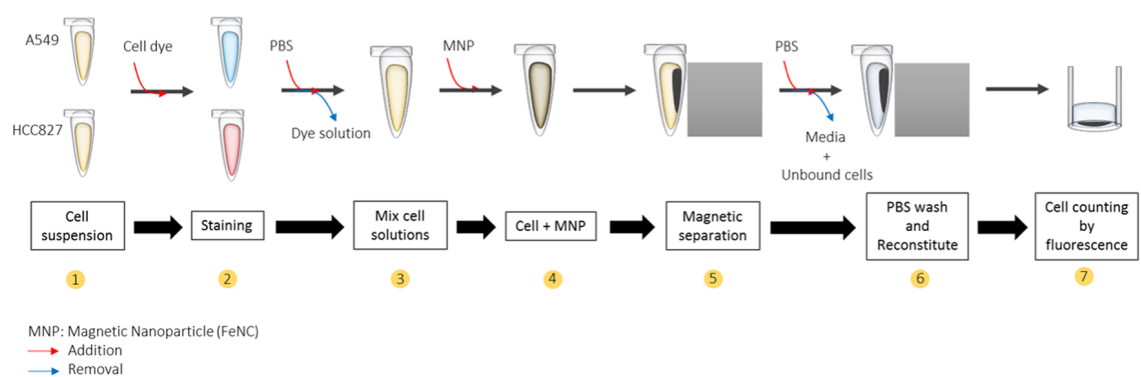

Figure 4. Steps in the immunomagnetic FeNC-mediated capture and isolation of cancer cells.

To isolate a maximum number of cells, we optimized two crucial parameters: antibody coating on FeNC and particulate concentrations of FeNC required for isolation. For evaluating the first parameter, we used FeNC with varying degrees of Ab on the surface and treated with the sample containing 100 cells of each cell line (Figure 5). We anticipated that increased Ab

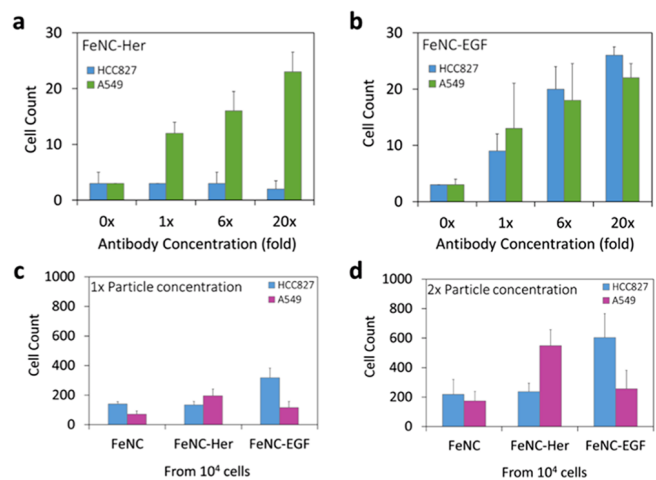

Figure 5. Capture selectivity and sensitivity of nanoparticles with increased (a, b) antibody coating on the surface and (c, d) particle concentrations on a 1:1 mixture of A549−HCC827 cells. Cells separated using immunomagnetic FeNC exhibit high specificity and sensitivity.

concentrations (1×, 6×, and 20× Ab) would increase the specificity of cell capture. That is, FeNC conjugated with a high concentration of herceptin (anti-Her2) should capture A549 cells with a higher degree of specificity. Similarly, FeNC conjugated with high concentrations of cetuximab (anti-EGFR) should capture both A549 and HCC827 with no preference for either one. As shown in Figure 5, the increase in Ab coating directly correlated with the number of cells captured. The data showed that the efficiency of FeNC−Her to capture A549 cells increased from 4- to 8-fold, while for FeNC−EGF to capture A549 or HCC827 cells increased from

4.3- to 7-fold or 3- to 9-fold, respectively. These results indicate the high selectivity of FeNC in capturing cancer cells based on the target receptor expression. For evaluating the second parameter, we treated different concentrations of FeNC−Her or FeNC−EGF or FeNC (control) with a cell mixture of $10^4$ cells per cell line, and the results are presented in Figure 5. In this experiment, we performed cell capture with varying amounts of NPs, and the results are presented in Figure 5. The data for FeNC−Ab showed a proportional increase in capture (2- to 3-fold) with an increase in particle concentration. The data also showed a 3-fold increase in capture selectivity for FeNC−Ab constructs as compared to that for the FeNC control at higher concentrations. The results indicated that a minimum concentration of $2 \times 10^6$ particles/mL would be required to achieve high sensitivity in capturing cells without increasing nonspecific binding counts. Together, the data established the ability of FeNC−Ab to precisely capture cancer cells based on their target expression.

**Magnetic Nanocubes for Capturing Epithelial Cancer Cells in Serum.** As a next step, we evaluated the ability of antibody-conjugated FeNC constructs to capture cancer cells spiked in serum to simulate clinical sampling conditions. In this study, we used FeNC (control), FeNC−Her, and FeNC−EGF for selective capturing of A549 and HCC827 cells independently. The cells were spiked in serum isolated from pig blood. Both FeNC−Her and FeNC−EGF successfully captured cells according to the receptor expression levels in A549 and HCC827 (Figure 6). As expected, the FeNC control showed negligible capture (Figure 6). Our results showed that FeNC−Her was highly specific and efficient in capturing A549 cells than HCC827 cells as compared to FeNC−EGF (92 vs 55%), while FeNC−EGF was more active in detecting HCC827 cells. The fold-change between FeNC−Her and FeNC−EGF in capturing A549 cells was about 3-fold, while the specificity of FeNC−EGF over FeNC−Her in capturing HCC827 was almost 4-fold. These data suggested that FeNC−Her was an ideal candidate for capturing A549, while FeNC−EGF was a candidate to capture HCC827 cells. Additionally, the data indicated that magnetic nanocubes were stable in capturing cells in serum and exhibit similar specificity and selectivity as that of aqueous solutions. Such studies with the Her/EGFR-targeted capture based on heterogeneous expression in cells have been previously demonstrated.[45]

Subsequently, we conducted a specificity test for FeNC−Her and FeNC−EGF in capturing positive target cells relative to control from a cell mixture in serum. For this experiment, we used samples spiked with an A549−HCC827 1:1 mixture (from $10^2$ to $10^4$ cells per cell line), wherein A549 was a

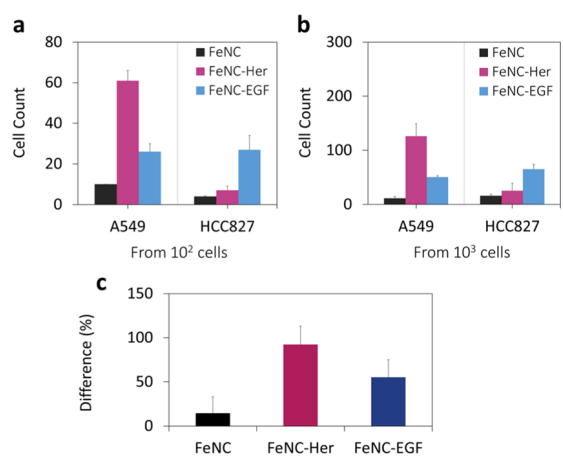

Figure 6. Demonstration of (a, b) the receptor-based capture using $10^2$ or $10^3$ cells and (c) construct specificity to A549 vs HCC827 in serum solution.

positive control for FeNC−Her, and HCC827 was a negative control. Similarly, we tested a sample spiked with HCC827−MCF-7 mixture (from $10^2$ to $10^4$ cells per cell line), wherein HCC827 was a positive control for FeNC−EGF, and MCF-7 was a negative control. Our results indicated a successful capture and isolation of cells by the constructs over their controls (Figure 7). Data showed a specificity difference of 9-

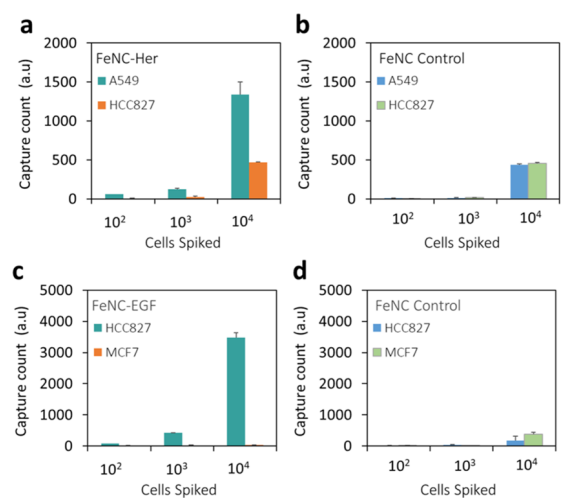

Figure 7. Specificity capture tests between (a, b) FeNC−Her and FeNC (nonspecific capture) for A549−HCC827 mixed spiked samples and (c, d) FeNC−EGF and FeNC (nonspecific capture) for HCC827−MCF-7 mixed spiked samples.

and 6-folds for FeNC−Her and FeNC−EGF, respectively, in capturing positive cells instead of negative control cells at the lowest cell spike. In both cases, the nonspecific capture by FeNC (control) was insignificant (<2%) compared to that of FeNC−Ab constructs. Similar results in a previous study with magnetic liposomes showed higher specificity when EGFR was targeted compared to EpCAM.[46] Another group showed that targeting Her2 and EGFR can complement and enhance epithelial marker-based targeting methods.[47] In summary, the data demonstrated the ability of FeNC to capture cancer cells based on the antibody−receptor specificity to oncogenic surface receptors such as Her2 or EGFR in serum.

**Magnetic Nanocubes for Capturing Mesenchymal Cancer Cells in Serum.** To test the ability of FeNC−Her and FeNC−EGF in capturing mesenchymal A549 and HCC827 cells, we spiked the EMT-induced cells in serum (along with negative HCC827 or MCF-7 epithelial cells) and analyzed the capture counts (Figure 8). We followed the fluorescent labeling strategy as described above to count the cells. FeNC−Her showed excellent capturing ability with A549 cells before and after EMT induction (Figure 8). In fact, FeNC−Her showed higher efficiency in capturing most of the cells after EMT. Similarly, FeNC−EGF also showed remarkable capturing ability before and after EMT conversion of HCC827 cells. If the sample contains ≤10 cells, we observed background noise and more significant deviations in the capture. However, data was consistent within the range tested (10, 50, and 100 spiked cells per each type). The data established the capability of FeNC−Her and FeNC−EGF in capturing cancer cells irrespective of their EMT status. These results are in agreement with a previous study where targeting mesenchymal cells by EMT-marker was 2.1-fold efficient than targeting epithelial markers.[48] In another study, the magnetic beads targeting EGFR were found to be 80% more efficient than targeting epithelial markers in a low EpCAM expressing cell line.[49] Together, the data strengthens the use of growth receptors such as Her2 or EGFR to capture mesenchymal cells.

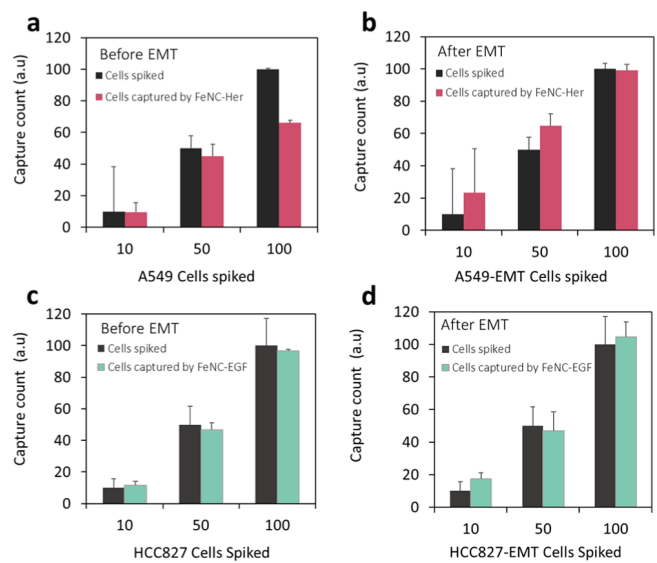

Figure 8. Cell capture before and after induction of EMT in (a, b) A549−HCC827 mixture and (c, d) HCC827−MCF-7 mixture by FeNC−Her and FeNC−EGF, respectively. EMT in A549 and HCC827 cells were stimulated, respectively.

**Biomarker Characterization of Captured Cells.** As a final step, we investigated the surface epithelial marker status of the captured cells. We performed fluorescent immunostaining using an anti-CK antibody (tagged with Alexa Flour 647) in the epithelial and mesenchymal captured cells (Figure 9a). Then, we analyzed the difference between the total cells captured and the cells that showed positive-CK staining (Figure 9b,c). A549 cells captured by FeNC−Her showed a 62.5% decrease in cells expressing CK after EMT stimulation ($p < 0.047$). Similarly, HCC827 cells captured by FeNC−EGF showed a 75% decrease in cells expressing CK after induction of EMT ($p < 0.005$). The drastic decrease in epithelial marker expression in cells confirmed that the captured cells are of mesenchymal type. Further, we showed that magnetic nanocubes capture and allow the detection of cytokeratin status in the captured cells. The ability to characterize biomarkers in captured cells is an essential step in developing technologies for clinical applications.[50−53] In summary, our results demonstrate that FeNC−Her and FeNC−EGF constructs could successfully capture, isolate, and allow the detection of epithelial marker status in captured cells.

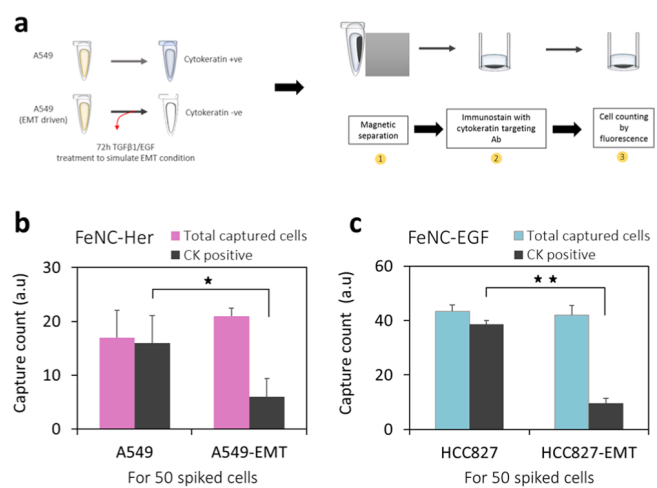

**Figure 9.** (a) Steps for EMT stimulation and magnetic separation; and staining and detection of cytokeratin in cells captured by (b) FeNC−Her and (c) FeNC−EGF.

## ■ CONCLUSIONS

We developed a novel magnetic nanocube-based tumor-cell capture technique for separating epithelial and/or mesenchymal cells. In this study, we have shown that surface markers such as Her2 and EGFR are unaffected by the EMT state and are potential biomarkers for capturing CTCs. The magnetic nanocubes can help in isolating specific cancer cells based on their marker and allow the detection of EMT and non-EMT population subsets within a given pool of cells. The results of this study underscore the importance of using growth receptors as a secondary target for capturing CTCs. As the tumor is always heterogeneous, the proposed method provides an opportunity to capture CTCs that express variable biomarkers. The results present in this study can aid in developing next-generation techniques for CTC isolation in clinics.

## ■ METHODS

**Materials.** We purchased iron(III) acetylacetonate, oleic acid, benzyl ether, poly(maleic anhydride-*alt*-1-octadecene) (PMAOD), *N*-(3-dimethylaminopropyl)-*N*′-ethylcarbodiimide hydrochloride (EDC), *N*-hydroxy sulfosuccinimide sodium salt (sulfo-NHS), 2-(*N*-morpholino)ethane sulfonic acid (MES), bovine serum albumin (BSA), MS-SAFE protease and phosphatase inhibitor, Triton X-100, Tween-20, and phosphate-buffered saline (PBS) from Sigma-Aldrich. Hexane, toluene, and chloroform were purchased from Acros Organics. Sodium hydroxide (NaOH), acetone, sucrose, sodium borate buffer, Hoechst dye, CellMask deep red membrane dye, Pierce ECL Western blotting substrate, Pierce BCA protein assay kit, and 3,3′,5,5′-tetramethylbenzidine (TMB) were purchased from Thermo Fisher Scientific. β-Actin, vimentin, EpCAM, cytokeratin primary antibody kit (7, 8, 17, 18, 19, PAN), EGFR, Her2, and both antimouse and antirabbit secondary horseradish peroxidase (HRP)-linked immunoglobulin G (IgG) antibodies were purchased from Cell Signaling. Carboxylic-poly(ethylene glycol)-amine was custom-synthesized from Nanocs and Laysan Bio. 10× Tris-buffered saline (TBS), 10× Tris/glycine/sodium dodecyl sulfate (SDS) buffer, 4−15% 10-well 50 μL ready Mini-Protean TGX, Precision Plus Protein dual-color standard ladder, and supported nitrocellulose membrane (0.2 μm) were purchased from Bio-Rad. Recombinant human EGF (carrier-free) and Alexa Fluor 647 tagged anticytokeratin PAN antibody were purchased from BioLegend. Recombinant human transforming growth factor β-1(TGFβ-1) and EGF were purchased from Novoprotein. Amicon 0.5 mL of 100 kDa regenerated cellulose ultracel centrifugal filter units were purchased from Merck Millipore Ltd. Cetuximab (Eli Lilly and Company) and herceptin (Genetech Inc.) were purchased from the hospital pharmacy.

**Instrumentation.** High-resolution JEOL transmission electron microscope (TEM) was used to measure and record the images of nanoparticles. The hydrodynamic size and ζ potential of the nanoparticles were measured using a Malvern Zetasizer Nano-ZS (Malvern Panalytical). The centrifugation was performed on a 5424 Eppendorf and refrigerated RC 6+ Sorvall centrifuge (Thermo Fisher Scientific). The pH was measured using a Seven Compact Mettler Toledo pH meter equipped with an InLab Micro electrode. We performed fluorescence imaging, fluorescence, and UV−vis absorption measurements using a Cytation 5 cell imaging multi-mode reader (BioTek Instruments Inc.). Gel electrophoresis was performed on a Bio-Rad Mini-PROTEAN Tetra system, and blots were transferred using a Genscript e-blot transfer system. We performed Western blot imaging and acquisition using Image Lab version 5.2.1 software on a ChemiDoc XRS system from Bio-Rad. Solvent extraction was performed on a Büchi rotavapor R-124 attached to a water bath and water-cooled distillation column. Finally, SQUID measurements were performed on a Quantum Design MPMS 3 magnetometer (Quantum Design).

**Synthesis of Iron Oxide Nanocubes.** We synthesized iron oxide nanocubes (FeNCs) using a high-temperature polyol reduction process in an organic phase. Briefly, a modified protocol based on previous studies was used for synthesizing particles.[34,54] In this procedure, iron(III) acetylacetonate (7.06 g, 20 mmol) was added to a mixture of oleic acid (1.129 g, 12.69 mL) and benzyl ether (104 g, 99.71 mL) in a three-neck 200 mL round-bottom flask (RBF) attached to a reflux condenser and a Schlenk line. This solution was degassed using Argon (20 kPa) for 1 h under constant stirring (400 rpm), and the reaction was kept under argon throughout the process. After initial degassing, the reaction temperature was increased to 295 °C at a heating rate of 20 °C/min under constant stirring (500 rpm). The reaction color turned jet black from reddish brown after 30 min, and at this stage, the solution temperature decreased to 100 °C. Next, the solution was then quickly cooled on ice and a mixture (80 mL) of hexane, toluene, and acetone with a ratio of 1:1:2 was added. This solution was stirred and sonicated to remove particles stuck to the stir bar. The solution was then transferred to a 500 mL beaker with a seal and allowed to precipitate for 15 h. The top solution was decanted, and the remaining solution containing the precipitate was centrifuged at 20 000*g* for 1 h (15 °C). The pellet was then resuspended and washed several times using a mixture of hexane and toluene (ratio of 1:1) until a slight clear supernatant formed. For each wash, the tube was kept over a neodymium magnet (pull force of 250 lb) for 15 min followed by the slow removal of the supernatant using a 10 mL pipette. This solution was then concentrated in hexane and washed five times (15 000g for 15 min) until a clear supernatant formed. Finally, the particles (20 mg yield) were resuspended in 4 mL of hexane and stored at room temperature.

**Hydrophilic Conversion of Iron Oxide Nanocubes.** The iron nanocubes synthesized using the polyol process resulted in the formation of hydrophobic particles. To convert them to hydrophilic particles for use in aqueous and physiological solutions, we replaced the organic coating with a functional amphiphilic polymer by modifying a previously published protocol.[36] Briefly, synthesized FeNCs (in hexane) were injected into a rapidly stirring (500 rpm) poly(maleic anhydride-alt-1-octadecene) (1.2 g, average molecular weight 30 000–50 000) solution dissolved in chloroform (100 mL) in a one-neck 200 mL RBF. After 2 h of stirring, the solvent was removed using a rotary evaporator (100 rpm, 55 °C) attached to a standard vacuum line protected by a liquid nitrogen cool trap, until a wet sludge was formed. The sludge was quickly mixed with 50 mL of 0.1 M NaOH solution and dispersed by sonicating for 1 h. The solution was then stirred (500 rpm) for 15 h at room temperature to hydrolyze and open the maleic anhydride rings. The solution was washed twice with 0.1 M NaOH solution at 20 000g (15 °C) for 1 h, followed by washing twice with a three-layer sucrose density gradient centrifugation (30, 20, and 10% sucrose). The pellet was resuspended and concentrated in 4 mL of 10 mM MES buffer and washed once. As a next step, 100 kDa centrifugal filters were equilibrated with MES buffer and used for removing excess polymer (6000g for 6 min). The solution was then again subject to a two-layer sucrose density gradient centrifugation (30 and 10% sucrose) at 15 000g for 15 min, followed by six washes with DI water. The FeNC pellet was then dispersed in 5 mL water and stored at room temperature.

**Synthesis of PEGylated Nanocubes.** To enhance the stability and solubility in aqueous solutions, the carboxyl groups of hydrophilic particles were activated using EDC/sulfo-NHS and conjugated with $NH_2$-PEG-COOH. Briefly, the particles were dissolved in 0.1 M MES buffer (pH 4.5) containing 2 mg EDC and 2 mg sulfo-NHS and kept for shaking (850 rpm) for 1 h at 28 °C. The particles were then centrifuged (15 000g for 15 min), and the pellet was dispersed in a 50 mM sodium borate buffer containing 10 mg carboxylic-poly(ethylene glycol)-amine (COOH-PEG-$NH_2$, MW 2000; 5 μmol) and 5 mg EDC (26 μmol). This solution was kept shaking (850 rpm) for 15 h at 22 °C. After the reaction, PEGylated FeNCs were washed with water two times (15 000g for 15 min), suspended in DI water, and stored at room temperature. PEGylated FeNCs were used for conjugation with antibodies.

**Antibody Conjugation.** We conjugated the antibody (Ab) to the hydrophilic PEGylated FeNCs by the traditional EDC/sulfo-NHS cross-linking procedure. Briefly, 100 μL (2 × 10[6] particles/mL) of FeNCs was washed with 0.1 M MES buffer (pH 4.5) prior to being activated. The pellet was dispersed in 0.1 M MES buffer (400 μL; pH 4.5) with 3 mg EDC (16 μmol) and 4 mg sulfo-NHS (18 μmol) at 28 °C for 3.5 h (shaking at 800 rpm). After activation, the particles were centrifuged (15 000g for 15 min), dispersed in 100 μL 1× PBS, and mixed with the antibody solution of either 4 μL of globin–Ab (10 mg/mL), 200 μL of herceptin (2 mg/mL), or 200 μL of cetuximab (2 mg/mL). The reaction was kept in a shaker (800 rpm) for 15 h at 22 °C and, upon completion, washed twice with 1× PBS (15 000g for 15 min). The final solution was dispersed in 1 mL of 1× PBS and stored at 4 °C. The FeNC–Ab–globin was used for estimating the protein conjugation efficiency using the known globin protein, while FeNC conjugated to cetuximab and herceptin was used to target EGFR and Her2 in cell studies. These particles were labeled as FeNC–EGF and FeNC–Her, respectively. Due to the ready availability of globin–Ab, we were able to estimate the conjugation efficiency using ELISA and anticipate that the conjugation efficiency of antibodies of EGF and Her to FeNC would be similar to that of globin–Ab to FeNC.

**Characterization of Magnetic Nanocubes.** We characterized FeNC using TEM and dynamic light scattering (DLS) technique. For TEM, 8 μL of 1/10th diluted FeNCs was placed over a CF300-Cu carbon film-based copper grid and air-dried for 10 min in a hot air oven (40 °C). TEM imaging for polymer-coated and PEGylated FeNCs showed a uniform size of 20 nm. For DLS measurements, 1 mL of 1/10th diluted FeNC solution was placed in a disposable UV-grade cuvette and analyzed. The magnetic properties were further investigated using the superconducting quantum interference device (SQUID) analysis. For this analysis, 5 mg of the lyophilized FeNC power was loaded in sealed SQUID sample holders. The magnetic moment was recorded using the temperature sweeping mode under zero-field-cooled (ZFC) conditions from 2 to 300 K at 1000 Oe. The corrected magnetic moment was then determined by a previously published procedure.[55]

**ELISA.** To determine the conjugation efficiency of the antibody bound to FeNC, we used an indirect enzyme-linked immunosorbent assay (ELISA). For this experiment, the FeNC–Ab–globin construct was used to evaluate the specificity to globin antigen; the reaction-supernatant was used to quantify the amount of conjugated globin antibody. PEGylated FeNCs were used as the negative control, and PBS was considered as the blank. Briefly, 100 μL of globin solution (0.1, 1, 10, and 100 μg/mL, respectively) was added to wells of a flat-bottom sterile 96-well plate and left undisturbed for 15 h at 4 °C. The wells were then washed with a wash buffer (0.5% Tween-20) followed by blocking with 200 μL of 2% BSA solution for 30 min at room temperature. As a next step, the wells were washed with a wash buffer prior to the addition of samples (100 μL). After 2 h, at 37 °C, the sample solutions were washed with a wash buffer. Subsequently, 50 μL of the 0.32 μg/mL secondary antibody (goat antirabbit IgG-HRP) was then added to the wells and incubated for 30 min at 37 °C followed by two washes with a wash buffer. Finally, 50 μL of the substrate solution (TMB peroxidase) was added to the wells and the reaction was stopped by adding 0.1 M HCl solution. The plates were then read by a Cytation 5 absorbance plate reader at 450 nm, and the absorption intensity was correlated to the concentration of the globin antibody conjugated to FeNC.

**SQUID Magnetic Analysis.** We evaluated the magnetic properties of FeNCs using superconducting quantum interference device (SQUID) equipped with a 7 T magnet. In this process, Josephson junctions were used to measure a change in the magnetic flux using a coil with a known inductance. For the magnetometry analysis, 5 mg of pure lyophilized FeNCs was loaded in a VSM sample holder. Measurements were made using a five-quadrant magnetic moment with varying external magnetic field analysis ($H_{max}$ of 5 T at $T$ = 2 K). Zero-field-cooled magnetic susceptibility and field-cooled warming measurements were performed using a sweep mode from 2 to 300 K at 1000 Oe. Using this analysis, the blocking temperature was estimated. Raw data was then corrected for sample shape and radial offsets and graphs were plotted.

**Isolation of Blood Serum.** Pig serum was obtained from an investigator at the University of Missouri for the spiking studies. Protocols in accordance with the ethical animal-handling regulations as approved by the University of Missouri were followed. Five milliliters of blood was collected from pigs in a sealed test tube. The clotted blood was centrifuged ($100g$ for 10 min at 4 °C), and the serum was isolated and stored at 4 °C. The serum was used for cancer cell spiking experiments.

**Cell Culture.** A549, HCC827, and MCF-7 human cancer cell lines (ATCC) were grown in the Roswell Park Memorial Institute (RPMI) 1640 medium (obtained from Gibco BRL). The media was supplemented with 4.5 g/L D-glucose, 25 mM $N$-(2-hydroxyethyl)piperazine-$N'$-ethane sulfonic acid (Hepes), 0.11 g/L sodium pyruvate, 1.5 g/L sodium bicarbonate, 2 mM L-glutamine, 10% heat-inactivated fetal bovine serum (FBS; Atlanta Biologicals), and 1% penicillin/streptomycin antibiotic solution. The cells were cultured in a humidified atmosphere of 95% air and 5% $CO_2$ at 37 °C (Thermo Scientific).

**EMT Conversion.** To induce epithelial-to-mesenchymal transition (EMT) artificially in cells, 120 ng of TGF$\beta$ (20 ng/mL) and 60 ng of EGF (10 ng/mL) protein were mixed in 6 mL serum-free media and treated with A549 cells. Treatment was performed at 60% confluence in a T25 flask for 72 h in a $CO_2$ incubator. Flasks were monitored every 24 h using a bright-field microscope, and images were obtained. At the end of the conversion, the flask was washed once with fresh RPMI media containing 10% FBS and twice with cold 1× PBS. The cells were then dislodged and used for magnetic capture.

**Capture and Detection of Cancer Cells.** To capture and detect spiked cancer cells, we developed a robust magnetic extraction process. In this experiment, the cells selected for capture and detection were either A549 (Her2+++, EGFR++) or HCC827 (EGFR+++), while control cells were either HCC827 (EGFR+++) or MCF-7 (Her2+++), respectively. For capturing the cells of interest, A549 or HCC827 cells with or without EMT treatment (induced by TGF$\beta$-1 and EGF) were dislodged from the T25 flask (70% confluence) and used for spiking experiment.[42,43] For all experiments, target cells such as A549 or HCC827 were live-stained red, while control cells such as HCC827 or MCF-7 were live-stained blue. The addition of control cells allows us to test the specificity of the construct in each study. Additionally, all capture experiments relied on live cells, while cytokeratin detection was performed on fixed cells. Briefly, the cells were suspended in 1× PBS and incubated with 40 $\mu$L of Hoechst (blue) dye or 40 $\mu$L of CellMask (deep red stain) for 20 min at 37 °C. The cells were washed with 1× PBS twice to remove excess dye (2000 rpm for 5 min). The cells were counted, and cell stock solution was prepared ($10^2/10^3$ or $10^4$ cells). Two hundred microliters of cell suspension ($10^2/10^3$ or $10^4$ cells) was mixed with 200 $\mu$L of control cells ($10^2/10^3$ or $10^4$ cells, respectively) and placed in a tube. Equal amounts of each target and control cell were also placed in separate tubes as nontreated control. For testing fixed concentration, 200 $\mu$L of the antibody-functionalized FeNC construct (FeNC–Her or FeNC–EGF; 2 × $10^6$ particles/mL) was added to the tube with cells and incubated at 37 °C for 2 h (gently shaken every 5 min). For evaluating varying concentrations, either the surface–antibody concentration on nanoparticles (1× (0.4 mg/reaction), 6×, and 20×) or nanoparticle concentration (1 × $10^6$ or 2 × $10^6$ particles/mL) was changed. Following the treatment, the tubes were placed (0.1 cm spacing) next to an N50-grade neodymium block magnet (with a surface field of 4833 G, a dipole moment (m) of 75.6 A m², and a permeance coeff. $P_c$ of 1.25) for 30 min. The supernatant was carefully removed, and the captured pellet was washed twice with 500 $\mu$L of 1× PBS using magnetic separation. The FeNC–Ab–cell solution was then resuspended in 200 $\mu$L of 1× PBS and transferred to a flat-bottom 96-well plate. To prevent any agglomeration, a hydrophobic low retention 200 $\mu$L tip was used to quickly mix and transfer without immersing the tip fully in solution. Cell counting of stained cells was performed using an automated fluorescence-based cell counter using the Cytation 5 imaging plate reader.

**Western Blot.** To perform Western blotting (WB), A549, HCC827, or MCF-7 cells (1 × $10^6$) with or without EMT treatment at 70% confluence were seeded in six-well plates. The cells were treated in serum-free media for a period of 72 h followed by whole-cell lysate preparation using lysis buffer containing 1× phosphatase inhibitor cocktail. Proteins were separated by 4−15% polyacrylamide gel electrophoresis (PAGE) and were transferred onto a nitrocellulose membrane for protein blot analysis. Membranes were washed with 1× TBST, blocked with a 5% BSA blocking buffer containing 0.5% normal goat serum (NGS) for 2 h, and incubated with a primary antibody overnight on a shaker at 4 °C. Blots were then washed before and after incubation with a secondary antibody and developed with the chemiluminescence system. Densitometry analysis was performed using Image Studio Lite software (LI-COR Biosciences).

**Detection of Cytokeratin.** To detect epithelial marker status, the cells captured by antibody-functionalized FeNC constructs (FeNC–Her or FeNC–EGF) were stained (Hoechst dye) and fixed by adding 500 $\mu$L of 4% paraformaldehyde for another 10 min at 37 °C. Next, 500 $\mu$L of 0.1% Triton X-100 was added and incubated for 5 min after staining and fixing. The rest of the steps were the same for the FeNC treatment. After cell capture and washing once with 1× PBS, 500 $\mu$L of 5% FBS solution in 1× PBS was added as a blocking agent for 30 min. After the removal of a blocking agent, 400 $\mu$L of Alexa Fluor 647 anticytokeratin antibody (2 $\mu$g/mL) solution diluted in 1× PBS cells was added to the tube and incubated for 2 h at room temperature. The cells were then washed twice with 1× PBS and then resuspended in 200 $\mu$L of 1× PBS. This solution was then quickly transferred to a 96-well plate for imaging and fluorescence-based cell counting using the Cytation 5 imaging plate reader.

**Statistics.** All statistics were performed using Minitab. The average values from triplicates were calculated and used for statistical analysis. Statistical significance was evaluated using Student's $t$ test, and $p < 0.05$ was considered significant.

## ■ ASSOCIATED CONTENT

### ⓈSupporting Information

The Supporting Information is available free of charge at https://pubs.acs.org/doi/10.1021/acsomega.0c02699.

    ELISA plot for FeNC–Ab–globin; TEM images of FeNC in different stages; DLS and $\zeta$ potential of FeNC;

and microscopy images of untreated A549 cells for a period of 0−72 h (Figures S1−S4) (PDF)

## AUTHOR INFORMATION

**Corresponding Author**

Raghuraman Kannan − *Department of Bioengineering and Department of Radiology, University of Missouri, Columbia, Missouri 65212, United States;* orcid.org/0000-0003-1980-3797; Email: kannanr@health.missouri.edu

**Authors**

Dhananjay Suresh − *Department of Bioengineering, University of Missouri, Columbia, Missouri 65212, United States;* orcid.org/0000-0003-3490-5636

Shreya Ghoshdastidar − *Department of Bioengineering, University of Missouri, Columbia, Missouri 65212, United States*

Abilash Gangula − *Department of Radiology, University of Missouri, Columbia, Missouri 65212, United States*

Soumavo Mukherjee − *Department of Bioengineering, University of Missouri, Columbia, Missouri 65212, United States*

Anandhi Upendran − *Department of Medical Pharmacology & Physiology and Institute of Clinical and Translational Science, University of Missouri, Columbia, Missouri 65212, United States*

Complete contact information is available at:
https://pubs.acs.org/10.1021/acsomega.0c02699

**Author Contributions**

⊥D.S. and S.G. contributed equally to this work. D.S. formed the initial hypothesis and detection method. D.S. and R.K. conceived the experiments. D.S. and S.G. synthesized and characterized the samples. D.S., S.G., and A.G. handled the antibody staining and magnetic separation. D.S. and S.G. handled the cellular imaging. D.S., S.M., and S.G. carried out the Western blotting. D.S. and A.U. handled the blood and serum isolation. D.S. performed data analysis and S.G. carried out the statistics. D.S. and R.K. wrote the manuscript.

**Notes**

The authors declare no competing financial interest.

## ACKNOWLEDGMENTS

R.K. acknowledges Michael J and Sharon R Bukstein Endowment for financial support of this work.